\documentclass[english,aps,prstper,onecolumn,showpacs,longbibliography,superscriptaddress,nofootinbib]{revtex4-2}
\usepackage[T1]{fontenc}	
\usepackage{geometry}		
\usepackage{graphicx}
\usepackage{epsfig}
\usepackage[above,below]{placeins}	
\usepackage{times}
\usepackage{csquotes} 
\usepackage{gensymb} 
\usepackage{enumitem}          
\setlist{nosep} 
\usepackage{url}
\usepackage{hyperref}  
\hypersetup{colorlinks=true,urlcolor=blue,citecolor=blue,linkcolor=blue}  

\begin{document}
\title{Breaking Barriers: Investigating Gender Dynamics in Introductory Physics Lab Classes}

 \author{Bilas Paul}
\email[Please address correspondence to ]{palb@farmingdale.edu}
  \affiliation{SUNY Farmingdale State College, Farmingdale, NY, 11735} 
\author{Shantanu Chakraborty}
  \affiliation{Valdosta State University, Valdosta, GA, 31698} 
 \author{Ganga Sharma}
  \affiliation{Fairmont State University, Fairmont, WV, 26554}

\begin{abstract}
\label{abstract}
\noindent 
The persistent underrepresentation of women and other gender minorities in physical science fields has been an ongoing concern. This study investigates gender dynamics in introductory physics laboratory courses, focusing on whether students of different gender identities exhibit equal inclination and confidence in conducting lab experiments, and whether they face barriers that impact their participation. Conducted across three institutions and involving non-physics STEM students enrolled in algebra-based and calculus-based physics courses, the study found mixed results, with two institutions showing no significant gender-based differences in participation levels during lab activities, while one institution demonstrated significant differences. Chi-square tests revealed no significant association between gender and task preference or comfortability, though the small dataset suggests the need for further investigation. While quantitative analysis provided limited evidence of systematic barriers, qualitative feedback revealed that some female students experienced challenges related to gender dynamics, such as perceived assumptions about competence, being overlooked during discussions, and hesitation to voice opinions in male-dominated groups. These findings highlight the complex influence of gender and institutional factors on laboratory experiences and underscore the need for creating inclusive environments that promote equitable engagement and participation for all gender identities in STEM education.

\end{abstract}

\maketitle

\section{INTRODUCTION }
\label{intro}
Women remain significantly underrepresented in physics and other STEM (science, technology, engineering, and mathematics) fields, particularly in academic settings ~\cite{AIPstat}. Despite significant efforts to increase diversity and inclusion, the disparity between male and female participation in STEM disciplines remains a persistent issue~\cite{Henderson2018AnEO}. This underrepresentation is influenced by various factors, including societal norms, stereotypes, and institutional barriers. Studies have examined the impact of gender stereotypes on female students’ performance in introductory physics courses, which often serve as foundational requirements for many STEM majors, highlighting potential barriers to their success~\cite{10.1103/physrevphyseducres.14.020119}. Additionally, comparisons of self-efficacy between male and female students with similar academic backgrounds have shed light on how self-perception can influence performance and persistence in physics~\cite{10.1103/physrevphyseducres.14.020123}.  

Studies have also scrutinized female students’ experiences in physics classes, particularly during lab experiments, to identify existing barriers. Investigations have examined issues related to equipment handling based on gender and race in both remote and in-person lab settings~\cite{10.1119/perc.2022.pr.dew}. Interestingly, some studies suggest that female students often take on leadership roles in group settings in physics labs to ensure tasks are completed and the group stays on track~\cite{10.1088/1361-6404/abd597}. However, gendered task divisions in labs can lead to disruptions in learning and equitable participation for female students~\cite{10.1088/1361-6404/ab7831}. Gender differences in perception and agency in physics labs have also been studied to determine if there are disparities in outcomes between men and women~\cite{10.1119/perc.2020.pr.kalende}. It has been noted that without active intervention to structure equitable group dynamics, gendered divisions of roles can emerge, potentially leading to inequitable learning experiences~\cite{PhysRevPhysEducRes.16.010129}.   

This study examines gender dynamics in introductory physics courses at mediumsized 4-year institutions, focusing on non-physics STEM majors who are required to take these courses as part of their degree programs. By investigating the performance, preferences, and potential barriers encountered by female students in these settings, the study aims to deepen understanding of the factors contributing to the underrepresentation of women in physics and STEM fields at large. Insights from this study are critical for promoting gender equity and inclusivity in STEM education and addressing the broader issue of women’s underrepresentation in these fields~\cite{PhysRevPhysEducRes.20.010102, PhysRevPhysEducRes.18.010102}.

\section{MATERIALS \& METHODS}
This study utilized a mixed-methods approach to investigate gender dynamics within introductory physics laboratory classes, with a particular emphasis on students’ engagement in conducting lab experiments. Participants consisted of non-physics majors enrolled in algebra-based and calculus-based introductory physics courses at three distinct institutions: Fairmont State University in Fairmont, West Virginia; SUNY Farmingdale State College in Farmingdale, New York; and Valdosta State University in Valdosta, Georgia (Table I).  

To provide context for the study, we summarize the demographics of the participating institutions. At SUNY Farmingdale, the racial composition includes 44.1\% White, 27.2\% Hispanic/Latino, 12.1\% Asian, 10.4\% Black or African American, and smaller proportions of other racial and ethnic groups, with over half of the student population identifying as minorities and 28.9\% classified as underrepresented minorities. The gender distribution at Farmingdale is 59.0\% male and 41.0\% female. Fairmont State University has a predominantly White student body (86.1\%), with smaller proportions of Black or African American (4.66\%), and even smaller representations from other racial and ethnic groups. Gender representation at Fairmont is 43.6\% male and 56.4\% female. At Valdosta State University, the student population comprises 44.0\% White, 37.0\% Black, 11.0\% Hispanic, and smaller proportions of other racial/ethnic groups, with 66.7\% female and 33.3\% male students. These demographic details provide a diverse backdrop for examining gender dynamics in laboratory settings.

\begin{table*}[htb!] 
  \caption{Summary of the experiments we analyzed, including the number of enroll students and their genders. Non-binary students were excluded from this analysis.}
  \label{tab:coding_scheme}
  \begin{ruledtabular}
    \begin{tabular}{ccccccc}
    \textbf{Institution} & \textbf{Course}& \textbf{Lab Experiments} &\textbf{Enrolled} & \textbf{Male} & \textbf{Female}& \textbf{Non-binary}\\
&&&\textbf{ students}&&&\\
      \hline
Fairmont State University & Algebra-based Physics II & Resistors \& Ohm's Law, Capacitors & 33& 20 &11 & 2 \\
&& and \(B\) Field in a  Current Carrying Wire &&&&\\
\hline
Farmingdale State College & Algebra-based Physics II & Gas Laws, Heat Engine,  & 24 & 13& 10& 1\\
&& and  Mapping Electric Potential &&&&\\
\hline
Valdosta State University & Calculus-based Physics II & Oscilloscope, Ohm's Law, & 28 & 19 & 8 & 1\\
&& and   Kirchhoff's Rule &&&&\\
    \end{tabular}\label{tab:summary_participant}
  \end{ruledtabular}
\end{table*}

Ethical considerations were addressed through institutional review board approval, ensuring compliance with ethical guidelines for research involving human participants. The anonymity and confidentiality of the participants were maintained throughout the data collection and analysis processes. The study’s limitations included potential observer bias during data collection and the reliance on self-reported survey responses. 

The subject pool for data collection comprised students primarily from aviation, biology, engineering, engineering technology, and pre-medical programs. These students were essential in providing a diverse and representative sample for the study, allowing for a comprehensive analysis of gender dynamics within the laboratory setting. 

Physics II was chosen as the focus of this study because students enrolled in Physics II have already completed the prerequisite introductory course, Physics I. This allowed us to assume that some initial barriers and unfamiliarity with physics concepts or laboratory equipment were mitigated. By focusing on Physics II, we could examine gender dynamics in a setting where students had foundational knowledge and prior exposure to physics lab environments, enabling a more nuanced analysis of engagement and participation. 

In the laboratory portion of these courses, students conducted experiments weekly in groups of three or four members. The research methodology encompassed both observational and survey-based approaches to capture a holistic view of students’ engagement levels and perspectives on participating in lab activities. Observations were conducted in-person and in real-time by the primary instructors during these lab sessions to assess student engagement and participation levels. Each instructor employed a cyclic observation method, observing student groups at approximately 20-minute intervals and  recording whether each student was actively involved in one of the following tasks: setting up equipment, data collection, note-taking, calculations, plotting graphs, or engaging in group discussions related to the experiment. The observational procedures were carefully integrated into the natural flow of lab activities to ensure minimal disruption and preserve typical student behavior. Based on their involvement, the time each student spent on various tasks was recorded for each experiment. This data was collected for three separate experiments (Table 1) for each laboratory group. These experiments were carefully chosen to cover essential topics in Physics II, ensuring they addressed key concepts students must understand for success in these courses. With a structure aimed at maintaining active student involvement throughout nearly the entire two-hour lab session, they provided ample opportunities for meaningful observation of students’ roles and interactions. For analysis, the time each student spent on various tasks across the three experiments was averaged, which provided an overall engagement profile. To visualize the differences in engagement levels, these average time distributions were then plotted as a histogram. This graphical representation allows for a clear comparison of student engagement, highlighting any potential gender differences in participation levels. To determine gender-based differences in engagement patterns between male and female students, independent samples t-tests were performed. These tests compared the average time distributions across gender groups, allowing us to identify any significant variations. This alignment with the study’s objectives enabled a focused investigation of participation dynamics.  

In addition to the observational data, a survey was administered to students to selfidentify their gender and indicate their preferred and most comfortable lab activities, choosing from options such as (A) setting up equipment, (B) data collection, (C) notetaking, calculations, and data visualization, and (D) report writing. Note-taking, calculations, and data visualization were grouped together as one category because these activities often occur simultaneously during data analysis and require similar cognitive processes~\cite{padilla2018decision, doi:10.1177/1473871611433713, }  of translating physical observations into mathematical and visual representations. To investigate potential associations between gender and both task preferences or comfort levels, chi-square tests were employed. These tests provided a statistical measure of whether the distribution of preferences and comfort levels differed significantly between gender groups. 

To identify potential barriers to full participation, the survey included open-ended questions asking students to describe any challenges they faced during lab activities. These questions were designed to capture experiences related to group dynamics, comfort levels with specific tasks, and any perceived biases or assumptions based on gender. This approach allowed students to express their views freely, providing qualitative insights into their experiences. These qualitative insights were analyzed thematically to identify recurring patterns and contextualize the broader findings of the study. 

Quantitative analysis involved calculating average observation scores, conducting statistical comparisons between genders using independent samples t-tests, and performing chi-square tests to examine potential associations between gender and both task preferences and comfortability during lab activities. Qualitative analysis entailed thematic examination of survey responses and qualitative insights derived from observational data to contextualize gender dynamics within the laboratory setting. 


\section{RESULTS}
Analysis of the time students spent on various laboratory tasks revealed varying patterns of student involvement by gender across the three institutions (Figure 1). While SUNY Farmingdale State College $(t=0.142, p=0.444)$ and Valdosta State University $(t=-0.475, p=0.319)$  showed no significant differences between male and female students' participation, Fairmont State University demonstrated significant gender differences $(t = 2.637, p = 0.007)$, as indicated by independent samples t-tests. This mixed finding suggests that gender-based participation patterns may be institution-specific.

\begin{figure*}[htb]
  \includegraphics[width=1\textwidth]{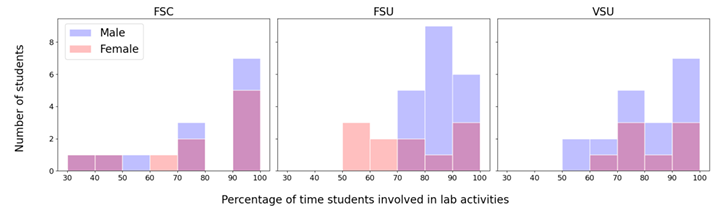}
 \caption{The distribution of time students spend on various tasks during physics lab sessions is shown across three institutions: SUNY Farmingdale State College (left), Fairmont State University (middle), and Valdosta State University (right). These tasks include setting up equipment, data collection, note-taking, calculations, plotting graphs, and participating in group discussions related to the experiments. The violet histograms represent male students, while the orange histograms represent female students.  \label{fig1}}
\end{figure*}

Building upon these initial findings of institution-specific differences in male and female student involvement, further analysis examined students' preferences for laboratory tasks (Figure 2) and comfort levels with physics laboratory activities (Figure 3). Chi-square tests were conducted separately for each institution to assess potential gender associations. The analysis revealed no significant gender associations with task preferences across all three institutions: SUNY Farmingdale State College $(\chi^2 = 1.583, p = 0.2083)$, Fairmont State University $(\chi^2 = 1.143, p = 0.2850)$, and Valdosta State University $(\chi^2 = 1.655, p = 0.1983)$. Similarly, comfort levels showed no significant gender associations at any institution: SUNY Farmingdale State College $(\chi^2 = 2.182, p = 0.1396)$, Fairmont State University $(\chi^2 = 0.172, p = 0.6783)$, and Valdosta State University $(\chi^2 = 1.772, p = 0.1831)$. These consistent findings across institutions suggest that both task preferences and comfort levels in laboratory activities were independent of gender in our dataset. 

\begin{figure*}[htb]
  \includegraphics[width=1\textwidth]{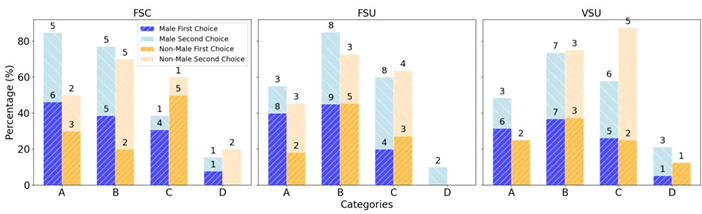}
 \caption{Distribution of students' preferences for various activities in physics lab sessions across three institutions: SUNY Farmingdale State College (left), Fairmont State University (middle), and Valdosta State University (right). The blue and orange bars represent the preferences of male and female students, respectively, with darker colors indicating first choice and lighter colors representing second choice. A = Equipment handling, B = Data collection, C = Note-taking, calculations, and plotting graphs, and D = Report writing. Some students indicated multiple first and/or second choices, while others had no first or second choice.   \label{fig2}}
\end{figure*}

\begin{figure*}[htb]
  \includegraphics[width=1\textwidth]{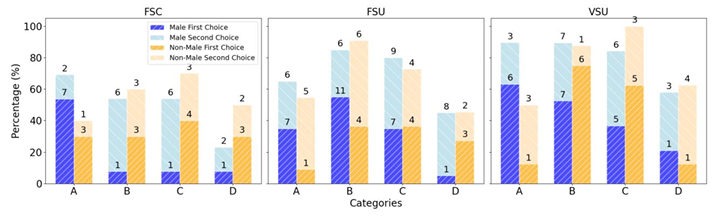}
 \caption{Distribution of students' comfort levels for various activities in physics lab sessions across three institutions: SUNY Farmingdale State College (left), Fairmont State University (middle), and Valdosta State University (right). The blue and orange bars represent the comfort levels of male and female students, respectively, with darker colors indicating first choice and lighter colors representing second choice. A = Equipment handling, B = Data collection, C = Notetaking, calculations, and plotting graphs, and D = Report writing. Some students indicated multiple first and/or second choices, while others had no first or second choice.  \label{fig3}}
\end{figure*}

While the majority of the quantitative data did not indicate gender-related barriers or challenges—such as disparities in participation, task preferences, or comfort levels between genders—in conducting physics lab experiments, qualitative feedback from a few students revealed that gender dynamics influenced their experiences in meaningful ways. Three key themes emerged from their comments: 

\paragraph{\textbf{Perceived Superiority}}
Some female students noted that male counterparts appeared to assume greater knowledge or competence, leading to imbalanced collaboration and contributions. As one student explained, "Sometimes as a woman, male partners just assume that they know more and do not want to mess up steps." This observation aligns with previous studies highlighting how gendered assumptions about competence can affect group dynamics and individual participation in STEM settings~\cite{10.1119/perc.2022.pr.dew, 10.1088/1361-6404/ab7831}. 

\paragraph{\textbf{Disregard for Contributions}}
In some instances, female students reported feeling overlooked or ignored by male team members, resulting in a diminished sense of belonging and engagement. One participant shared, "Being the only woman in a lab group with all men can be challenging at times. I am overlooked and not listened to, which affects my comfort level." Similar patterns of exclusion in collaborative settings have been noted as a barrier to equitable participation, particularly for women in male-dominated environments~\cite{PhysRevPhysEducRes.16.010129}. 

\paragraph{\textbf{Fear of Invalidation}}  
Certain participants expressed reluctance to share their opinions or correct mistakes due to fear of having their contributions dismissed based on gender. A student reflected, "Being in a lab group where I am the only girl made it difficult for me to feel comfortable enough to speak out on my opinion. I was often too nervous to try and correct a lab member mistake with the fear my lab member may discredit my finding or shrug off my input since I am just a girl." These sentiments echo findings that women in STEM often face challenges asserting their voices in group settings due to implicit biases or societal expectations~\cite{10.1103/physrevphyseducres.14.020123, 10.1088/1361-6404/abd597}. 

Although these experiences were reported by a minority of students, they highlight the subtle yet significant ways in which gender dynamics can impact participation, comfort, and engagement in laboratory settings. Sentiment analysis of the responses reveals a recurring sense of discomfort and marginalization, underscored by words like "nervous," "overlooked," and "challenging." 

These findings underscore the importance of addressing gender biases and fostering inclusive educational environments. Promoting open communication, equitable collaboration, and appreciation for diverse perspectives can help create a supportive atmosphere that encourages full participation and engagement for all students in STEM education. Continued efforts to develop inclusive pedagogical strategies, such as structured group roles and bias training, have shown promise in reducing gender disparities in STEM~\cite{PhysRevPhysEducRes.20.010102, PhysRevPhysEducRes.18.010102}.

\section{DISCUSSION}
This study investigated gender dynamics in introductory physics laboratory classes at medium-sized 4-year institutions, focusing on participation, task preferences, comfort levels, and potential barriers. The findings provide valuable insights into how gender influences laboratory experiences and highlight areas for promoting equity in physics education. 

The quantitative analysis of participation patterns showed varying results across institutions. While SUNY Farmingdale State College and Valdosta State University demonstrated no significant gender differences in participation ($t = 0.142, p = 0.444$ and $t = -0.475, p = 0.319$, respectively), Fairmont State University showed a significant difference ($t = 2.637, p = 0.007$). This institutional variation suggests that gender-based participation patterns may be influenced by local factors, such as institutional culture, classroom dynamics, or demographic composition. The significant result at Fairmont State University warrants further exploration to identify the underlying factors contributing to this variability. However, the small sample size in this study limits the generalizability of these findings, emphasizing the need for larger datasets to provide a more comprehensive understanding of gender dynamics in laboratory settings. 

Further analysis of students’ task preferences and comfort levels revealed no significant associations with gender at any of the three institutions. This suggests that male and female students reported similar preferences and comfort levels in laboratory activities. These findings indicate that while institutional factors may influence participation patterns, students' subjective experiences of task preferences and comfort were generally gender neutral. However, the absence of significant associations should be interpreted with caution given the modest sample size, which may not have been sufficient to detect subtle gender-based trends or variations across institutions. Expanding the dataset in future research could provide deeper insights into how gender shapes students’ interactions and engagement with specific laboratory tasks. 

While the quantitative findings provided limited evidence for systematic genderbased barriers, the qualitative results offered valuable insights into how gender dynamics can influence laboratory experiences. Female students reported challenges such as perceived assumptions of greater competence by male peers, being overlooked during group discussions, and hesitations to voice their opinions due to fear of dismissal. These observations are consistent with prior research documenting the impact of implicit biases and societal expectations on participation and collaboration in STEM settings~\cite{10.1119/perc.2022.pr.dew, 10.1088/1361-6404/abd597, 10.1088/1361-6404/ab7831, PhysRevPhysEducRes.16.010129, 10.1103/physrevphyseducres.14.020123}. Although these accounts were reported by a minority of students, they highlight subtle but significant ways in which gender dynamics can affect engagement, participation, and a sense of belonging in collaborative learning environments. Addressing these dynamics is critical for fostering equitable and inclusive STEM education~\cite{PhysRevPhysEducRes.20.010102, PhysRevPhysEducRes.18.010102}. 

The findings of this study underscore the need for targeted interventions to promote inclusivity in physics laboratory settings. Strategies such as implementing structured group roles, fostering open communication, and providing bias-awareness training for instructors can help mitigate the impact of implicit biases and create a more equitable learning environment~\cite{PhysRevPhysEducRes.20.010102, PhysRevPhysEducRes.16.010129}. Additionally, institutional efforts to cultivate a culture that values diverse perspectives and supports collaboration can further enhance student engagement and participation~\cite{PhysRevPhysEducRes.18.010102, 10.1119/perc.2020.pr.kalende}. These approaches align with broader efforts to reduce gender disparities in STEM and support the success of all students.

The study is ongoing, with data collection efforts aimed at expanding the dataset and enhancing the analysis. This research provides an opportunity to extend into other physical science and engineering disciplines to examine gender disparities and develop inclusive strategies for equitable participation and success for all students. Exploring these dynamics across various academic domains can offer insights into common challenges and aid in developing inclusive strategies to promote equitable participation and success for all students. 

\acknowledgments{}
The authors extend their gratitude to their students for their participation in this study. Author BP thanks the Provost's Office of SUNY Farmingdale State College for its support. Author SC thanks Valdosta State University Office of Sponsored Programs and Research Administration (OSPRA) for its support.  Author GS thanks Dr. Deb Hemler, Chair of Natural Science Department of Fairmont State University  for creating conducive research and learning environment within the department.

\bibliographystyle{unsrt}
\bibliography{main}

\end{document}